\documentclass[letterpaper, 10 pt, conference]{ieeeconf}  

\IEEEoverridecommandlockouts                              

\usepackage[noadjust]{cite}
\usepackage{graphicx,xcolor,epstopdf}
\usepackage{amsmath,amsfonts,amssymb}
\usepackage{stfloats}
\usepackage{booktabs}
\newcommand\cross[2][{}]{{{#2}_{#1}^\times}}
\usepackage[normalem]{ulem}

\definecolor{newcolor}{rgb}{0.0, 0.0, 0.0}

\usepackage{flushend}

\usepackage{hyperref}
\hypersetup{
	colorlinks,
	linkcolor={red!75!black},
	citecolor={purple!75!black},
	urlcolor={blue!50!black}
}

\newtheorem{theorem}{Theorem}

\newtheorem{remark}{Remark}

\newcommand{\R}[1]{\mathbb{R}^{#1}}
\newcommand{\so}[1]{\mathfrak{so}(#1)}
\newcommand{\SO}[1]{\mathsf{SO(#1)}}
\newcommand{\SE}[1]{\mathsf{SE(#1)}}
\renewcommand{\S}[1]{\mathsf{S}^{#1}}
\newcommand{\tr}[1]{\mathrm{tr}\left[#1\right]}

\newcommand{\LTD}[1]{\mathsf{T}^*_I \L_R\left(\mathsf{D}_R\,#1\right)}
\newcommand{\I}{\mathcal{I}}
\newcommand{\B}{\mathcal{B}}

\title{\LARGE \bf
Rigid-Body Attitude Control on $\SO{3}$ using Nonlinear Dynamic Inversion
}

\author{Hafiz Zeeshan Iqbal Khan, Farooq Aslam, Muhammad Farooq Haydar, and Jamshed Riaz
\thanks{$^{1}$Hafiz Zeeshan Iqbal Khan is with Centers of Excellence in Science and Applied Technologies, Islamabad, Pakistan, as well as the Institute of Space Technology, Islamabad, Pakistan. Farooq Aslam, Muhammad Farooq Haydar, and Jamshed Riaz are with the Institute of Space Technology, Islamabad, Pakistan.}
}

\begin{document}

\maketitle
\thispagestyle{empty}
\pagestyle{empty}

\begin{abstract}

This paper presents a cascaded control architecture, based on nonlinear dynamic inversion (NDI), for rigid body attitude control. The proposed controller works directly with the rotation matrix parameterization, that is, with elements of the Special Orthogonal Group $\SO{3}$, and avoids problems related to singularities and non-uniqueness which affect other commonly used attitude representations such as Euler angles, unit quaternions, modified Rodrigues parameters, etc. The proposed NDI-based controller is capable of imposing desired linear dynamics of any order for the outer attitude loop and the inner rate loop, and gives control designers the flexibility to choose higher-order dynamic compensators in both loops. In addition, sufficient conditions are presented in the form of linear matrix inequalities (LMIs) which ensure that the outer loop controller renders the attitude loop almost globally asymptotically stable (AGAS) and the rate loop globally asymptotically stable (GAS). Furthermore, the overall cascaded control architecture is shown to be AGAS in the case of attitude error regulation. Lastly, the proposed scheme is compared with an Euler angles-based NDI scheme from literature for a tracking problem involving agile maneuvering of a multicopter in a high-fidelity nonlinear simulation.

\end{abstract}

\section{Introduction}\label{sec:Intro}

The rigid-body attitude control problem is central to numerous aerospace and robotics applications. Given the highly nonlinear nature of the problem, control strategies based on feedback linearization have received considerable attention over the years \cite{snell1992nonlinear,enns1994dynamic,mokhtari2005robust,voos2009nonlinear,lee2009feedback}. In general, feedback linearization uses coordinate transformation and feedback to achieve exact cancellation of certain nonlinearities, thereby transforming a nonlinear dynamical system into a linear, or partially linear, dynamical system for which a suitable controller is then designed \cite{Slotine1991,Isidori1985}.

A feedback linearization method that has been studied extensively for aerospace applications is known as nonlinear dynamic inversion (NDI). Early results on NDI-based flight control, such as \cite{snell1992nonlinear,enns1994dynamic}, used Euler angles to describe rigid-body attitude. The same parametrization was also used to develop feedback linearizing control laws for quadrotor UAVs \cite{mokhtari2005robust,voos2009nonlinear,lee2009feedback}. However, in recent years, there has been growing interest in the design of NDI-based attitude control laws which work directly with the rotation matrix representation, also known as the direction cosine matrix (DCM).
These rotation matrices evolve on matrix Lie group $\SO{3}$, known
as Special Orthogonal Group. Since the rotation matrix provides an attitude representation which is both globally defined and unique \cite{Chaturvedi2011}, it can be used to develop attitude control laws which are plagued neither by the kinematic singularities associated with Euler angles nor the problem of unwinding associated with the unit quaternion attitude representation. Consequently, several researchers have sought to develop control schemes based on nonlinear dynamic inversion and incremental nonlinear dynamic inversion for attitude and position control \cite{caverly2016nonlinear, akhtar2020feedback, craig2020geometric, spitzer2021feedback}.
In \cite{caverly2016nonlinear}, dynamic inversion is used to control the attitude and airspeed of a high-altitude long-endurance flexible aircraft with the aircraft attitude being described using the rotation matrix parametrization. Input-output linearization is performed with the angular velocity $\omega$ and the aircraft forward velocity taken as the outputs. Thereafter, a geometric PID controller \cite{goodarzi2013geometric} is used for the attitude dynamics. In a similar vein, \cite{akhtar2020feedback} addresses the rigid-body attitude stabilization problem on $\SO{3}$ using partial state-feedback (or input-output) linearization. The authors investigate different output functions for obtaining locally and almost globally stabilizing feedback linearizing controllers with well-behaved zero dynamics. However, they do not consider the problem of attitude tracking or issues related to robustness. On the other hand, \cite{craig2020geometric} develops a feedback linearizing controller on the Special Euclidean Group $\SE{3}$ for attitude and position control of a quadrotor UAV operating in a windy environment. Dynamic inversion is used in conjunction with a geometric PD controller \cite{lee2010geometric} for the attitude dynamics and a variable-gain algorithm for handling rotor thrust saturation. Another effective solution has been developed in \cite{spitzer2021feedback}, where feedback linearization is used in conjunction with a learned acceleration error model to account for modeling errors and external disturbances, and to obtain a controller suitable for aggressive quadrotor flight.

As noted above, the feedback linearizing controllers developed in \cite{caverly2016nonlinear,akhtar2020feedback,craig2020geometric} utilize geometric PD or PID control laws such as those developed in \cite{goodarzi2013geometric,lee2010geometric}. It can be desirable from both a theoretical and a practical viewpoint to extend these geometric feedback linearization approaches so that they encompass a broader class of stabilizing linear dynamic controllers. To this end, this paper addresses the rigid-body attitude tracking problem on $\SO{3}$ using feedback linearization and linear dynamic compensation. In particular, a cascaded control architecture is considered which consists of an outer attitude loop and an inner angular rate loop with linear dynamic compensation in both the attitude and velocity loops. Sufficient conditions are obtained which ensure that the attitude loop is almost globally asymptotically stable (AGAS) and the rate loop globally asymptotically stable (GAS). \textcolor{newcolor}{These conditions are expressed in the form of linear matrix inequalities (LMIs). 
Furthermore, in the case of attitude regulation, we show that the overall cascaded architecture renders the closed-loop system to be AGAS.}

The main contribution of the paper is to extend existing geometric nonlinear control approaches so that they include more general linear dynamic controllers and provide practitioners with greater freedom in designing feedback linearizing attitude control laws on $\SO{3}$. Moreover, we hope that the developments detailed in this paper will allow control designers to combine linearization-based synthesis methods with geometric feedback linearizing control laws, as well as make it easier to incorporate linear models for actuator and/or sensor dynamics into the problem formulation. 
The rest of the paper is structured as follows: essential background and important results are summarized in Section \ref{sec:Prelim} along with some remarks on notation, and the main results are presented in Section \ref{sec:GeomtetricNDI}. Thereafter, Section \ref{sec:MulticopterExample} presents an example of agile maneuvering of a multicopter to demonstrate the effectiveness of the proposed scheme, and Section \ref{sec:Conc} concludes the discussion.

\section{Preliminaries}\label{sec:Prelim}
In this section, the rigid body attitude dynamics and kinematics are briefly discussed, and few key properties of associated operators are revisited from literature for completeness. Before presenting attitude dynamics of a rigid body, let us briefly introduce some notation. $\I$ is inertial frame, fixed and centered at earth’s surface. $\B$ is body fixed frame, centered at C.G. of the rigid body.\\
Let $\omega$ be the angular velocity of body $\B$ w.r.t. to $\I$ expressed in body frame $\B$. Then the rotational dynamics can be written as
\begin{equation}\label{eq:DynEqs}
  \dot{\omega} = J^{-1}\big[\tau - \omega\times J\omega - f(\omega,\mu)\big],
\end{equation}
where $J$ is the inertia matrix, $\tau$ is the control torque, and $f(\omega,\mu)$ contains other torques acting on the body e.g. aerodynamic damping, gravitational torques, etc. The parameters vector $\mu$, assumed to be either measured or estimated, is considered here to incorporate effects of other parameters such as aerodynamic angles, Mach number, dynamic pressure, etc.

The orientation $R$ of a rigid body, attitude transformation matrix from body frame $\B$ to inertial frame $\I$, evolves over $\SO{3}$, i.e. a Lie group containing all $3\times3$ orthogonal rotation matrices of determinant $+1$, commonly known as Special Orthogonal group and defined as:
\begin{equation}\label{eq:SO3}
\SO{3} \triangleq \big\{ R \in \R{3\times3} \mid R^\top R = RR^\top = I_3,\, \det(R) = 1 \big\}.
\end{equation}

Then the attitude kinematics of rigid body, also known as \emph{Poisson's Kinematical Equations (PKEs)} \cite{Stevens2015}, can be written as
\begin{equation}\label{eq:KinEqs}
  \dot{R} = R \cross{\omega},
\end{equation}
where if $\omega = [\omega_1,\omega_2,\omega_3]^\top\in\R{3}$, then
\begin{equation}\label{eq:HAT}
  \cross{\omega} \triangleq \begin{bmatrix}
                            0 & -\omega_3 & \omega_2 \\
                            \omega_3 & 0 & -\omega_1 \\
                            -\omega_2 & \omega_1 & 0
                          \end{bmatrix}.
\end{equation}

Here the \emph{cross} map $\cross{(\cdot)}:\R{3}\mapsto\so{3}$, transforms a vector in $\R{3}$  to its cross product form such that, $a\times b = \cross{a}b$ for any $a,b\in\R{3}$, where the \emph{Lie Algebra} $\so{3}$ is a vector space, or more precisely the tangent space of $\SO{3}$ at identity i.e. $\so{3} = \mathsf{T}_I\SO{3}$ and it can be written as follows: 
\begin{equation}\label{eq:so3}
\so{3} \triangleq \big\{ S \in \R{3\times3} \mid S^\top = -S \big\}.
\end{equation}

It is worth noting that the \emph{hat} map is an isomorphism. Its inverse is denoted by the \emph{vee} map $\vee : \so{3}\mapsto\R{3}$. Some important properties of the \emph{hat} map, which will be required in subsequent sections, are listed as follows \cite{lee2011geometric}:
\begin{subequations}\label{eq:PrelimProps}
\begin{align}
&\cross{x}y = x \times y = -y \times x = -\cross{y}x, \\
&\tr{A\cross{x}} = \frac{1}{2}\tr{\cross{x}(A-A^\top)} = -x^\top (A-A^\top)^\vee, \\
&\cross{x}A+A^\top\cross{x} = \left[ \left(\tr{A}I - A\right)x \right]^\times, \\
&R\cross{x}R^\top = (Rx)^\times, \\
&\tr{(\cross{x})^2} = -2x^\top x,
\end{align}
\end{subequations}
\noindent
for any $x,y\in\R{3}$, $A\in\R{3\times 3}$, and $R\in\SO{3}$. 

\section{Geometric NDI Control}\label{sec:GeomtetricNDI}
\begin{figure*}
  \centering
  \includegraphics[width=0.75\linewidth]{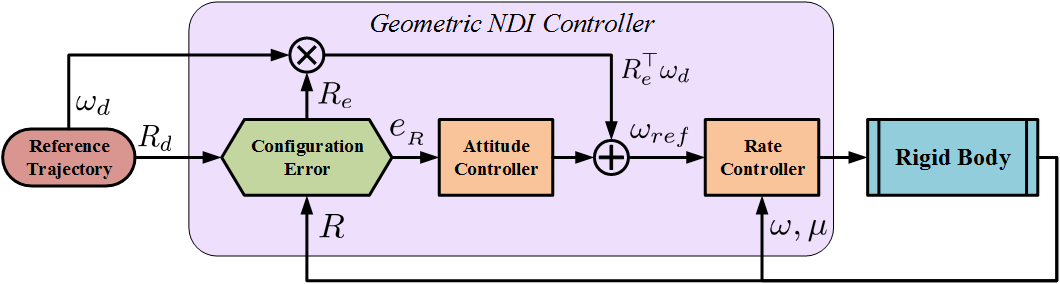}
  \caption{Geometric NDI Control Architecture}\label{fig:Blkdiag}
\end{figure*}

In this section the main results are presented. A two-loop geometric NDI structure for attitude control of a rigid body is proposed. The time scale separation is assumed between cascaded loops, and can be easily enforced by appropriate choice of controller gains. The complete control architecture is shown in Fig. \ref{fig:Blkdiag}. In particular it can be observed that the proposed architecture is similar to that of a standard NDI controller, except the geometric configuration error and a feed-forward term, which are precisely the components which renders the attitude loop almost globally asymptotically stable.


\subsection{NDI based Rate Control}
The nonlinear dynamic inversion based attitude rate control, presented in this section, is a slightly generalized version of the results presented in \cite{Khan2020}. This not only allows the user to choose a higher order compensator but also incorporate the feedback filters and feed-forward terms. Consider the following NDI control law:
\begin{equation}\label{eq:NDI_Rate}
\begin{split}
\dot{x}_\omega &= A_\omega x_\omega + B_\omega \omega + B_{\omega_{ref}} \omega_{ref}\\
\tau &= \omega\times J \omega + f(\omega,\mu) \\
&{}\quad+ J \left[C_\omega x_\omega +   D_\omega \omega + D_{\omega_{ref}} \omega_{ref} \right]
\end{split}
\end{equation}
\noindent
Substituting it in the dynamics \eqref{eq:DynEqs}, the closed-loop system can be written as,
\begin{equation}\label{eq:LEM1_003}
\begin{bmatrix} \dot{x}_\omega \\ \dot{\omega} \end{bmatrix} = \mathcal{A}_\omega \begin{bmatrix} x_\omega \\ \omega \end{bmatrix} + \begin{bmatrix} B_{\omega_{ref}}  \\ D_{\omega_{ref}}  \end{bmatrix} \omega_{ref}
\end{equation}
where,
\begin{equation}\label{eq:LEM1_001}
\mathcal{A}_\omega \triangleq \begin{bmatrix}
  A_\omega & B_\omega \\
  C_\omega & D_\omega
\end{bmatrix},
\end{equation}

Suppose the matrix $\mathcal{A}_\omega$ is Hurwitz which guarantees global asymptotic stability of \eqref{eq:DynEqs}. Moreover, by taking Laplace transform of \eqref{eq:LEM1_003} it can be easily shown that the control law \eqref{eq:NDI_Rate} induces the following desired dynamics,
\begin{equation}\label{eq:LEM1_002}
\omega(s) = [sI - \Gamma_\omega(s)]^{-1}\Gamma_{\omega_{ref}}(s) \omega_{ref}(s)
\end{equation}
where,
\begin{equation*}
\begin{split}
\Gamma_\omega(s) &\triangleq C_\omega [sI-A_\omega]^{-1}B_\omega + D_\omega \\
\Gamma_{\omega_{ref}}(s) &\triangleq C_\omega [sI-A_\omega]^{-1}B_{\omega_{ref}} + D_{\omega_{ref}}
\end{split}
\end{equation*}

\begin{remark}
For only proportional controller, $A_\omega$, $B_\omega$, and $C_\omega$ are empty matrices, $D_\omega = -K_\omega$, and $D_{\omega_{ref}} = K_\omega$. In that case only $K_\omega$ needs to be positive definite as in \cite{Khan2020}, and it enforces the first order desired dynamics i.e. $[sI+K_\omega]^{-1}K_\omega$.
\end{remark}
\subsection{Geometric NDI based Attitude Control}
Before proceeding with the development of geometric NDI controller, it is worth mentioning that the Lie group $\SO{3}$ is not a vector space, so it is not closed under addition operation. Therefore, let the desired attitude be $R_d$, and define the attitude error as
\begin{equation}\label{eq:Err}
  R_e \triangleq R_d^\top R.
\end{equation}
Where the matrix $R_e$ represents the attitude transformation from the body frame ($\B$) to the desired body frame ($\B_d$). Since the desired attitude $R_d$ also evolves on $\SO{3}$, therefore
\begin{equation}\label{eq:DesAttDyn}
  \dot{R}_d = R_d \omega_d^\times.
\end{equation}
Here it must be noted that this $\omega_{d}$ is different from $\omega_{ref}$ in previous subsection. Thus, the error dynamics can be written as,
\begin{equation}\label{eq:ErrDyn_Att}
  \dot{R}_e = R_e \cross[e]{\omega},
\end{equation}
where $\omega_e \triangleq \omega - R_e^\top\omega_d$.

Another important point is that for control design on non-Euclidean manifolds e.g. $\SO{3}$,  a notion of norm of a point $R$ is required, or more precisely the manifold needs to be equipped with a \emph{Riemannian} metric to define several geometric notions like length, angle etc. For attitude control using rotation matrices, commonly used metrics include the chordal and geodesic metrics \cite{Chaturvedi2011,Bullo2019}, as well as a metric recently proposed in \cite{lee2012exponential} for improved performance (relative to the chordal metric) in the case of large-angle rotational errors. In this work only chordal metric is considered, since it will result in a smooth control law, in contrast to others, which result in discontinuous controllers. The Chordal metric is defined as follows,
\begin{equation}\label{eq:Metric}
  \langle R_a , R_b \rangle_c \triangleq \left\|I - R_a^\top R_b\right\|^2_F = 2\,\tr{I - R_a^\top R_b}
\end{equation}
for any $R_a,R_b\in\SO{3}$, where $\|\cdot\|_F$ represents \emph{Frobenius} norm. Now the configuration error function can be defined as,
\begin{equation}\label{eq:ConfErrFunc}
\Psi(R_d,R) \triangleq \frac{1}{4}\langle R_d , R \rangle_c = \frac{1}{2}\tr{I - R_e}
\end{equation}
Then according to \cite{lee2011geometric} attitude error vector ($e_R$) is the left-trivialized derivative of the configuration error function, so
\begin{equation}\label{eq:AttErr}
e_R \triangleq \LTD{\Psi(R_d,R)} = \frac{1}{2}\left(R_e - R_e^\top\right)^\vee
\end{equation}
There are many metrics for $\SO{3}$ are available in literature e.g. Geodesic (shortest-path) \cite{Bullo2019}, Chordal \cite{Chaturvedi2011} etc. A succinct overview of metrics on $\SO{3}$ is available in \cite{berkane2017hybrid}.
Applying \emph{hat} map on \eqref{eq:AttErr}, taking its derivative and using the identities \eqref{eq:PrelimProps}, results in $2\cross[R]{\dot{e}} = \left[\left(\tr{R_e}I - R_e^\top\right) \omega_e\right]^\times$. Thus, the derivative of attitude error can be written as,
\begin{equation}\label{eq:AttErrDer}
\dot{e}_R = \mathcal{E}(R_e) \omega_e
\end{equation}
where $\mathcal{E}(R_e) \triangleq \frac{1}{2}\left(\tr{R_e}I - R_e^\top\right)$. It is worth noting that these dynamics cannot be inverted directly because $\mathcal{E}(R_e)$ is not invertible at an attitude error of $\exp\left(\pm \frac{\pi}{2}s\right)$ and $\exp\left(\pm \pi s\right)$ for any $s\in\S{2}$.



Now to obtain local attitude error dynamics, consider the Euler-axis parametrization. Under the small error angle assumption, the attitude error can be written as
\begin{equation}\label{eq:ApproxRE}
R_e \approx I + e_\Phi^\times
\end{equation}
where $e_\Phi = \Phi - \Phi_d$, and $\Phi = [\phi,\theta,\psi]^\top$ and $\Phi_d = [\phi_d,\theta_d,\psi_d]^\top$ are the actual and desired Euler angles, respectively. Thus using \eqref{eq:AttErr}, attitude error vector ($e_R$) can be approximated as
\begin{equation}\label{eq:ApproxErr}
  e_R \approx e_\Phi,
\end{equation}
Now using \eqref{eq:ApproxRE} in \eqref{eq:ErrDyn_Att}, and after some simplification, we get,
\begin{equation}\label{eq:LinErrDynTemp}
\cross[\Phi]{\dot{e}} \approx \left[I + \cross[\Phi]{e}\right]\cross[e]{\omega}
\end{equation}
Moreover, under the small error assumption, the product terms in \eqref{eq:LinErrDynTemp} would be negligible, therefore, local linearized error dynamics can be written as,
\begin{equation}\label{eq:LocalErrDyn}
  \dot{e}_\Phi \approx \omega_e
\end{equation}

\begin{theorem}[Attitude Controller]
Consider the following NDI control law:
\begin{equation}\label{eq:NDI_Att}
\begin{split}
\dot{x}_R &= A_R x_R + B_R e_R\\
\omega  &= R_e^\top\omega_d + C_R x_R + D_R e_R
\end{split}
\end{equation}
Suppose there exist a positive definite matrix $P$ such that
\begin{equation}\label{eq:LEM2_001}
\mathcal{Q} \triangleq \begin{bmatrix}
D_R & \star \\
PB_R + \frac{1}{2}C_R^\top & A_R^\top P + PA_R
\end{bmatrix} \prec 0.
\end{equation}
then the control law \eqref{eq:NDI_Att}:
\begin{enumerate}
  \item renders the desired attitude ($R_e = I$) to be the almost globally asymptotically stable equilibrium of \eqref{eq:ErrDyn_Att},
  \item gives exact local tracking performance ($\Phi(t) = \Phi_d(t)$),
  \item induces the following local desired dynamics about the stable equilibrium, if feedforward term ($R_e^\top\omega_d$) is ignored in control law
\begin{equation}\label{eq:LEM2_002}
\Phi(s) = -[sI - \Gamma_\Phi(s)]^{-1}\Gamma_\Phi(s) \Phi_d(s)
\end{equation}
where $\Gamma_\Phi(s) = C_R\left[sI - A_R\right]^{-1}B_R + D_R$.
\end{enumerate}
\end{theorem}
\begin{proof}
To prove first statement, considering the Lyapunov function as $\mathcal{V} = 2 \Psi(R_d,R) + x_R^T P x_R$, it can be seen directly from \eqref{eq:ConfErrFunc} that $\mathcal{V}$ is positive definite and radially unbounded. Moreover, its derivative can be computed as follows,
\begin{equation*}\label{eq:LEM2_003}
\begin{split}
\dot{\mathcal{V}} =&\,- \tr{\dot{R}_e} + \dot{x}_R^T P x_R + x_R^T P \dot{x}_R \\
 =&\,e_R^T \left(\omega -R_e^T\omega_d\right)  + x_R^T (A_R^T P + P A_R) x_R \\
     &\;{}+ x_R^T P B e_R + e_R^T B^T P x_R \\
=&\,e_R^TD_Re_R+\frac{1}{2}\left(e_R^T C_R x_R + x_R^T C_R^T e_R\right) \\
 &\;{}+ x_R^T (A_R^T P + P A_R) x_R + x_R^T P B e_R + e_R^T B^T P x_R  \\
 =& \begin{bmatrix} e_R^T  &  x_R^T \end{bmatrix} \mathcal{Q} \begin{bmatrix} e_R  \\  x_R \end{bmatrix} < 0.
\end{split}
\end{equation*}
Therefore, the control law \eqref{eq:NDI_Att} drives the configuration error function ($\Psi$) to zero. The critical points of $\Psi$ are the solutions $R_e\in\SO{3}$ to the equation $\Psi = 0$ or $\tr{I-R_e} = 0$, which are given by $R_e = I$ (desired equilibrium), and $R_e = \exp(\pm\pi\cross{s})$ (undesired equilibria) for any $s\in\S{2}$ \cite{Bullo2019}. However, using Chetaev's instability theorem \cite[Theorem 4.3]{Khalil2002}, it can be
shown that the undesired equilibria are unstable (for details see \cite{lee2010geometric}). Thus the desired equilibrium is almost globally asymptotically stable.

Proof of second and third statements is straightforward; substituting Eqs. \eqref{eq:NDI_Att}, \eqref{eq:ApproxRE} and \eqref{eq:ApproxErr} in Eq. \eqref{eq:LocalErrDyn}, and taking Laplace transform of resulting local linear closed-loop as $\Phi(s)=\Phi_d(s)$. If the feedforward term ($R_e^\top\omega_d$) is ignored from control law, then small error assumption it can be approximated as $R_e^\top \omega_d \approx \dot{\Phi}_d$. This results in local error angle dynamics as $\dot{e}_{\Phi} = \omega - \dot{\Phi}_d$, which upon substituting Eqs. \eqref{eq:NDI_Att} and taking Laplace transform yields \eqref{eq:LEM2_002}.
\end{proof}

\begin{remark}
For only proportional controller, $A_R$, $B_R$, $C_R$, and therefore $P$ as well, are empty matrices and $D_R = -K_R$. In that case $K_R$ needs to be positive definite to ensure almost global asymptotic stability. It also enforces the first order local desired dynamics i.e. $[sI+K_R]^{-1}K_R$.
\end{remark}

\subsection{Stability Guarantees of Cascaded Architecture}
In this subsection we will discuss the stability of cascaded architecture for set-point tracking or regulation problems, i.e. ($\omega_d = 0$, $\dot{\omega}_d = 0$). Using Eqs. \eqref{eq:LEM1_003}, \eqref{eq:AttErrDer}, and \eqref{eq:NDI_Att} we can write complete cascaded closed loop error dynamics as,
\begin{equation}\label{eq:cLDyn01}
\begin{bmatrix}
\dot{e}_R \\ \dot{\omega}_e \\ \dot{x}_R \\ \dot{x}_\omega
\end{bmatrix} =
\begin{bmatrix} 0 & \mathcal{E}(R_e) & 0 & 0 \\
D_{\omega_{ref}} D_R & D_\omega & D_{\omega_{ref}} C_R & C_\omega\\
B_R & 0 & A_R & 0 \\
B_{\omega_{ref}} D_R & B_\omega & B_{\omega_{ref}} C_R & A_\omega
\end{bmatrix}
\begin{bmatrix}
e_R \\ \omega_e \\ x_R \\ x_\omega
\end{bmatrix},
\end{equation}
For simplicity lets denote $x_K = [x_R, x_\omega]^\top$, then we can write Eq. \eqref{eq:cLDyn01} as,
\begin{equation}\label{eq:cLDyn02}
\begin{bmatrix} \dot{e}_R \\ \dot{\omega}_e \\ \dot{x}_K \end{bmatrix} = \begin{bmatrix} 0 & \mathcal{E}(R_e) & 0  \\ A_{21} & A_{22} & A_{23}\\ A_{31} & A_{32} & A_{33} \end{bmatrix} \begin{bmatrix} e_R \\ \omega_e \\ x_K \end{bmatrix},
\end{equation}
where,
\begin{align*}
A_{21} &\triangleq D_{\omega_{ref}} D_R,                                  & A_{22} &\triangleq D_\omega,                                   & A_{23} &\triangleq \begin{bmatrix}D_{\omega_{ref}} C_R & C_\omega\end{bmatrix},             \\
A_{31} &\triangleq \begin{bmatrix}B_R\\B_{\omega_{ref}} D_R\end{bmatrix}, & A_{32} &\triangleq \begin{bmatrix}0\\B_{\omega} \end{bmatrix}, & A_{33} &\triangleq \begin{bmatrix}A_R & 0 \\ B_{\omega_{ref}} C_R & A_\omega\end{bmatrix}. 
\end{align*}
Now lets present the main stability results in following theorem.

\begin{theorem}
The cascaded closed loop system \eqref{eq:cLDyn02} is almost globally asymptotically stable for set-point tracking and regulation problems ($\omega_d = 0, \dot{\omega}_d = 0$), if there exits a positive scalar $p_{11}$, an scalar $p_{21}$, symmetric positive definite matrices $P_{22}$, and $P_{33}$, and a matrix $P_{32}$, such that following the LMIs hold.
\begin{subequations}\label{eq:TH2}
\begin{equation}\label{eq:TH2_01}
\mathcal{P} \triangleq \begin{bmatrix} p_{11} I & p_{12} I & 0 \\ \star & P_{22} & P_{23} \\ \star & \star & P_{33} \end{bmatrix} \succ 0
\end{equation}
\begin{equation}\label{eq:TH2_02}
\mathcal{M} \triangleq \begin{bmatrix} M_{11} & M_{12} & M_{13} \\
\star  & M_{22} & M_{23} \\
\star  & \star  & M_{33} \\
\end{bmatrix} \prec 0
\end{equation}
\end{subequations}
here the submatrices are defined as follows,
\[
\begin{split}
M_{11} &\triangleq p_{12}(A_{21} + A_{21}^\top),\\
M_{22} &\triangleq 2p_{12}I + P_{22}A_{22} + A_{22}^\top P_{22} + P_{23}A_{32} + A_{32}^\top P_{23}^\top,\\
M_{33} &\triangleq P_{23}^\top A_{23} + A_{23}^\top P_{23} + P_{33}A_{33} + A_{33}^\top P_{33},\\
M_{12} &\triangleq p_{11}I + p_{12}A_{22} + A_{21}^\top P_{22} + A_{31}^\top P_{23}^\top,\\
M_{13} &\triangleq p_{12}A_{23} + A_{21}^\top P_{23} + A_{31}^\top P_{33},\\
M_{23} &\triangleq P_{22}A_{23} + A_{22}^\top P_{23} + A_{32}^\top P_{33} + P_{23}A_{33}.
\end{split}
\]
\end{theorem}

\begin{proof}
Consider the lyapunov function,
\begin{equation}\label{eq:Lyap}
\begin{split}
\mathcal{V} = 2 p_{11} \Psi &+ \omega_e^\top P_{22} \omega_e + 2 p_{12}  e_R^\top \omega_e \\
&{}+ x_K^\top P_{33} x_K + 2 \omega_e^\top P_{23} x_K
\end{split}
\end{equation}
\noindent
Now using the fact that \cite[Proposition 1]{goodarzi2013geometric},
\begin{equation}
\Psi \geq \frac{1}{2} \|e_R\|^2.
\end{equation}

Therefore, it can be easily seen that \eqref{eq:TH2_01} ensures the positive definiteness and radial unboundedness of $\mathcal{V}$. So, the Lyapunov rate along trajectories of \eqref{eq:cLDyn02} can be written as follows:
\begin{equation*}
\begin{split}
\dot{\mathcal{V}} &= 2 p_{11} \dot{\Psi} + 2 \omega_e^\top P_{22} \dot{\omega}_e + 2p_{12} (e_R^\top\dot{\omega}_e + \omega_e^\top\dot{e}_R) \\
&\quad{}+ 2x_K^\top P_{33} \dot{x}_K + 2(\omega_e^\top P_{23} \dot{x}_K + x_K^\top P_{23} \dot{\omega}_e) \\
&= z^\top \mathcal{M} z + p_{12}\omega_e^\top \left(\mathcal{E}(R_e) + \mathcal{E}(R_e)^\top - 2I\right)\omega_e\\
\end{split}
\end{equation*}
Using the fact that the matrix ($\mathcal{E}(R_e) + \mathcal{E}(R_e)^\top - 2I$) is negative semi-definite for all $R_e \in \SO{3}$, we can bound lyapunov rate as follows
\begin{equation}\label{eq:Vdot}
\dot{\mathcal{V}} \le z^\top \mathcal{M} z  < 0
\end{equation}
Therefore, the cascaded control architecture drives the configuration error function ($\Psi$) to zero, alongwith $\omega_e$ and $x_K$. The critical points of $\Psi$ are the solutions $R_e\in\SO{3}$ to the equation $\Psi = 0$ or $\tr{I-R_e} = 0$, which are given by $R_e = I$ (desired equilibrium), and $R_e = \exp(\pm\pi\cross{s})$ (undesired equilibria) for any $s\in\S{2}$ \cite{Bullo2019}. However, using Chetaev's instability theorem \cite[Theorem 4.3]{Khalil2002}, it can be shown that the undesired equilibria are unstable (for details see \cite{lee2010geometric}). Thus the desired equilibrium ($R_e=I$, $\omega_e = 0$, $x_K=0$) is almost globally asymptotically stable. 
\end{proof}  
\vspace{-0.4cm}
\textcolor{newcolor}{\begin{remark}
It is worth noting that for tracking problems ($\omega_d \ne 0$), we need additional cancellation torque in rate-controller \eqref{eq:NDI_Rate}, more precisely ($R_e^\top \dot{\omega}_d - \cross[e]{\omega} R_e^\top \omega_d$). Furthermore, with this additional cancellation term along with the assumption of no gain and/or filter in rate loop feedback path (i.e. $B_{\omega_{ref}} = -B_\omega$ and $D_{\omega_{ref}} = -D_\omega$), it can be shown that the feasibility of LMIs \eqref{eq:TH2} also ensures AGAS for cascaded architecture for tracking problems. 
\end{remark}}


\section{Agile Maneuvering of a Multicopter: An Example}\label{sec:MulticopterExample}
\begin{figure}
  \centering
  \includegraphics[width=0.5\linewidth]{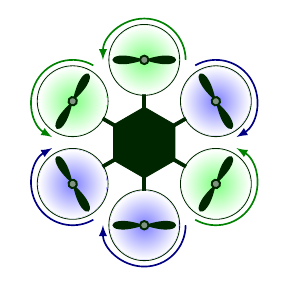}
  \caption{A Hexacopter Configuration}\label{fig:Hex}
\end{figure}

In this section, the agile maneuvering of a multicopter is considered to demonstrate the effectiveness of the proposed control scheme. Co-planar multicopters, as defined in \cite[Definition 2.1]{Khan2019}, are multicopters in which thrust vectors of all rotors are parallel in hover conditions for all control inputs. For such multicopters the function $f(\omega,\mu) = \kappa \omega$ in \eqref{eq:DynEqs}, where $\kappa$ is rotational damping coefficient. In this paper, an example of hexacopter is considered, see Fig. \ref{fig:Hex}. The proposed geometric NDI scheme is compared with an Euler angles based NDI control scheme proposed in \cite{Khan2020}. The parameters of hexacopter considered are available in \cite[Table I]{Khan2020}. A high fidelity nonlinear simulation is used, which also includes actuator dynamics and saturation limits on each motor RPMs, and to distribute the desired torques, \emph{Pseudo-Inverse} based control allocation scheme is used, for more details see \cite{Khan2020}.

A maneuver consisting of two flips (rotation of $720^\circ$) about roll axis followed by two flips about pitch axis, is considered, or more precisely,
\begin{equation}\label{eq:RefCmd}
\bar{R}_d(t) = \begin{cases}
        \exp(2\pi t \cross[1]{e}), & \mbox{if } 0 \leq t \leq 2 \\
        \exp(2\pi(t-2.5)\cross[2]{e}), & \mbox{if } 2.5 < t \leq 4.5 \\
        I, & \mbox{otherwise}.
      \end{cases}
\end{equation}
where $e_1 = [1,0,0]^T$ and $e_2 = [0,1,0]^T$. This maneuver is executed by generating a filtered reference $(R_d,\omega_d)$ using the second-order geometric filter developed in \cite[Section VI-C]{Invernizzi2020}. In particular, this filter is designed such that its linearized counterpart has a natural frequency of $15$ rad/s and a damping ratio of $0.707$. For comparison purpose, same desired dynamics and controller gains are used for the proposed scheme as for Euler angle based NDI presented in \cite{Khan2020}.

Though the presented approach can consider any desired dynamics, here a PD-type controller is used, or more precisely a lead compensator of the form $k_p + \frac{k_d s}{\tau_fs+1}$, for each channel of rate loop, with $k_p = 4.2$, $k_d = 0.42$, and $\tau_f = 10$. A first order lag filter at 100 Hz is used in feedback to mitigate sensor noise with a sensor delay of 5 ms in each channel. For analysis these pure delays are approximated by third order Pad\'{e} approximation. This gives an overall $12^\text{th}$ order rate loop control law \eqref{eq:NDI_Rate}. It can be seen that this ensures $\mathcal{A}_\omega$, as defined in \eqref{eq:LEM1_001}, to be Hurwitz. Moreover, for attitude loop, PID-type controller of the form $k_p + \frac{k_i}{s+\varepsilon} + \frac{k_d s}{\tau_fs+1}$ is used for each channel, with $k_p = -27.75$, $k_i = -1.85$, $k_d = -5.55$, $\varepsilon = 0.001$, and $\tau_f = 10$. It is worth noting that instead of a pure integrator and derivative, a lead-lag compensator is used, which makes the controller practically implementable. The feasibility of LMI \eqref{eq:LEM2_001} was checked by using a state-space realization of this controller, which was obtained by MATLAB's ``\texttt{ssdata}'' command. To ensure time scale separation, these controllers are designed on linearized models of each channel, and ratio of bandwidth of attitude loop to that of rate loop was kept higher than 4. With the selected gains this ratio was $7.580$, $6.121$, and $4.048$ for roll, pitch and yaw channels, respectively. \textcolor{newcolor}{Moreover, for stability of cascaded architecture, LMIs \eqref{eq:TH2} were checked to be feasible using YALMIP and SeDuMi toolboxes \cite{lofberg2004yalmip,sturm1999using}.}

\begin{figure}
  \centering
  \includegraphics[width=0.95\linewidth]{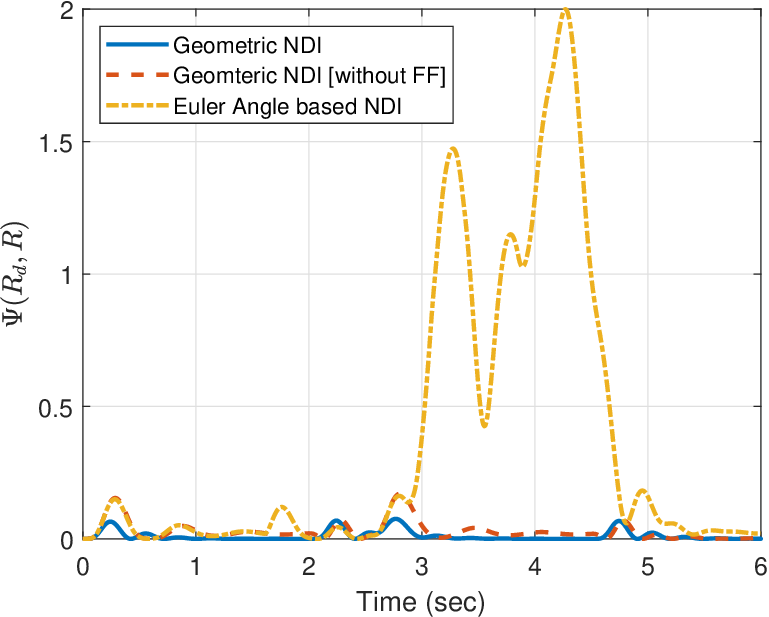}
  \caption{Configuration Error Function - $\Psi(R_d,R)$}\label{fig:PSI}
\end{figure}

\begin{figure}
  \centering
  \includegraphics[width=0.95\linewidth]{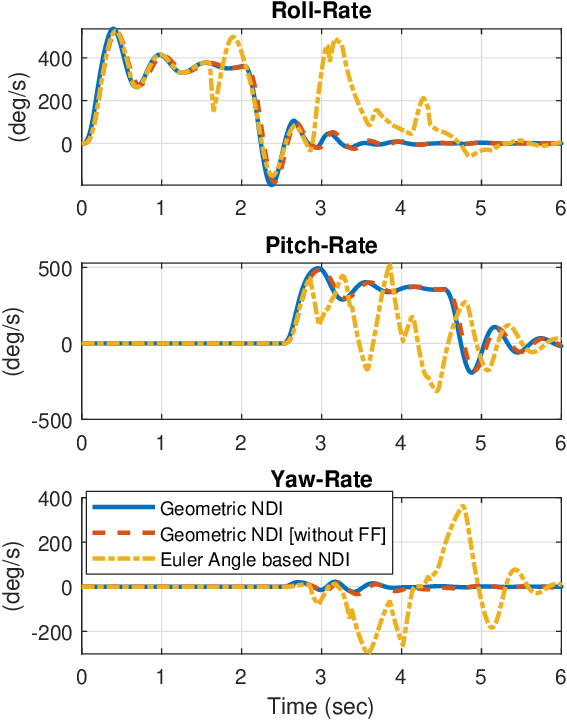}
  \caption{Body Rates - $\omega$}\label{fig:Rates}
\end{figure}

\begin{figure}
  \centering
  \includegraphics[width=0.95\linewidth]{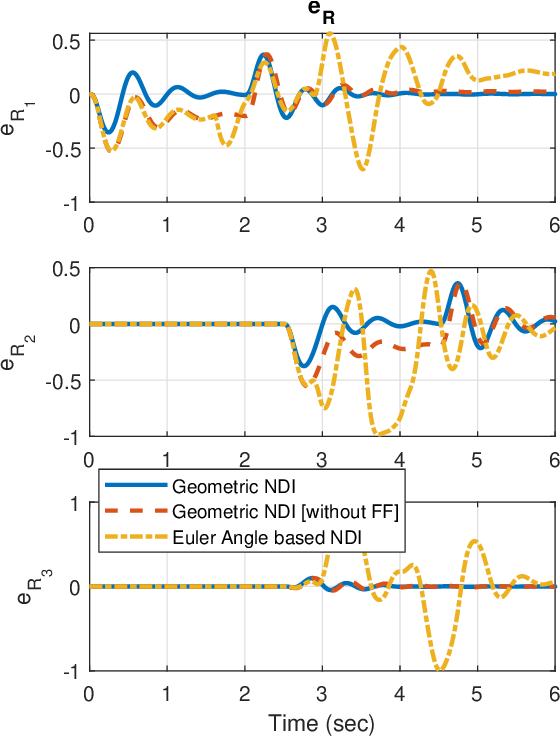}
  \caption{Attitude Error Vector - $e_R$}\label{fig:eR}
\end{figure}
\vfill
\begin{figure}
  \centering
  \includegraphics[width=0.95\linewidth]{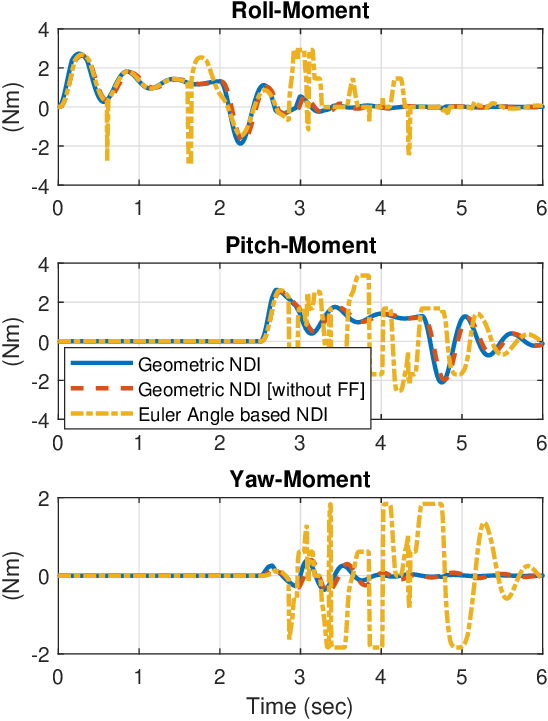}
  \caption{Control Effort - $\tau$}\label{fig:tau}
\end{figure}

Figure \ref{fig:PSI} shows the variation of the configuration error function ($\Psi(R_d,R)$), Fig. \ref{fig:Rates} shows the angular rates, Fig. \ref{fig:eR} shows the attitude error vector ($e_R$), and Fig. \ref{fig:tau} shows the control effort ($\tau$). It must be noted that Fig. \ref{fig:tau} shows the actual torque which is applied on the body, not the one demanded by the controller. Moreover, due to control allocation, actuator dynamics and saturation limits on motor RPMs, these demanded and actual torques are not necessarily equal. It can be easily seen that Euler angle based NDI controller barely survived the flips about roll axis with very degraded performance, and gets unstable during flips about pitch axis. However, the proposed geometric NDI scheme, both with and without feed-forward term ($R_e^\top \omega_d$), gives good performance during flips about both axes. Moreover, it can also be seen that the presence of feed-forward term significantly enhances the control performance at a cost of slightly larger control effort. 

\section{Conclusion}\label{sec:Conc}
In this work, a novel nonlinear dynamic inversion based cascaded control architecture is presented for the rigid body attitude control problem. The proposed control law uses the rotation matrix parameterization
\textcolor{newcolor}{and ensures almost global asymptotic stability in the case of attitude error regulation}. In particular, the proposed scheme is capable of enforcing desired linear dynamics of any order in both the attitude and velocity loops, and gives control designers the flexibility to use higher-order linear controllers in both loops while ensuring stability guarantees by just checking the feasibility of given LMIs. \textcolor{newcolor}{For practical applications, it is recommended that the inner velocity loop be at least three to five times faster than the outer attitude loop, as is standard practice in NDI-based cascaded control architectures.}



\bibliography{References}

\end{document}